\documentclass[aps,prb,superscriptaddress,twocolumn,no-footinbib,showpacs,floatfix,amsmath,amssymb]{revtex4-1}

\usepackage{graphicx}
\usepackage{dcolumn}
\usepackage{bm}
\usepackage[normalem]{ulem}
\usepackage{subfigure} 
\usepackage{soul,xcolor}
\usepackage{siunitx}
\usepackage{textcomp}

\begin{document}

\title{Band Gaps and Wavefunctions of Electrons Coupled to Pseudo Electromagnetic Waves in Rippled Graphene}

\author{Ramon Carrillo-Bastos}   
\affiliation{Facultad de Ciencias, Universidad Aut\'{o}noma de Baja California, Apdo. Postal 1880, 22800 Ensenada, Baja California, M\'{e}xico.}
\author{Gerardo G. Naumis}
\email{naumis@fisica.unam.mx}
\affiliation{Depto. de Sistemas Complejos, Instituto de F\'{i}sica, Universidad Nacional Aut\'{o}noma de M\'{e}xico (UNAM). Apdo. Postal 20-364, 01000 M\'{e}xico D.F., M\'{e}xico}

\begin{abstract}
The effects of a propagating sinusoidal out-of-plane flexural deformation  in the electronic properties of a tense membrane of graphene are considered within a non-perturbative approach, leading to an electron-ripple coupling. The deformation is taken into account by introducing its corresponding pseudo-vector and pseudo-scalar potentials in the Dirac equation. By using a transformation to the time-cone of the strain wave, the Dirac equation is reduced to an ordinary second-order differential Matthieu equation, i.e., to a parametric pendulum, giving a spectrum of bands and gaps determined by resonance conditions between the electron and ripple wave-vector (G), and their incidence angles. The location of the nth gap is thus determined by $E\approx n v_{F}\hbar G$, where $v_F$ is the Fermi velocity. Physically, gaps are produced by diffraction of electrons in phase with the wave. The propagation is mainly in the direction of the ripple.  In the case of a pure pseudoelectric field and for energies lower than a certain threshold, we found a different kind of equation. Its analytical solutions are in excellent agreement with the numerical solutions.  The wavefunctions can be expressed  in terms of the Matthieu cosine and sine functions, and for the case of a pure pseudo-electric potential, as a combination of Bessel functions.
\end{abstract}

\maketitle

\section{Introduction}
Out-of-plane acoustic modes are characteristic vibrations in graphene. For low frequency they are easiest to be excited and the ones that carry most of the vibrational energy\cite{jiang2015review}. They generate local dynamical bond stretching, bending and twisting\cite{prb-jiang}. Bond stretching or strain is by far the most important for electrons, since it causes greater impact on the tunneling parameter\cite{neto2009electronic}. In general, lattice deformations can be expressed in the low energy Hamiltonian by a gauge field\cite{PRL-Vozmediano-92,PRL-Kane-Mele-97,vozmediano2010gauge}. In particular, when a modest value of strain is applied to graphene\cite{vozmediano2010gauge,naumis-review}, the change in distance between carbon atoms can be mapped  with the inclusion of the  scalar deformation potential\cite{suzuura2002phonons} and pseudo-gauge field\cite{vozmediano2010gauge}; although four other terms are allowed by symmetry considerations\cite{Amorim}. It has been argued that screening can lead to strong suppression of the deformation potential\cite{Park-2014,Sohier-prb-2014} for static strain, but even when partially screened it affects the electronic properties of graphene\cite{barraza-prb}. On the other hand, the magnetic-like effects of the pseudo-gauge field are strong and measurable\cite{Amorim,naumis-review}; experimental results of Levy et. al. \cite{levy2010strain} confirmed the formation of pseudo Landau levels\cite{ll-theory} in strained graphene with values of magnetic field as high as 300 teslas. Moreover, the coupling between the pseudo-gauge field and the pseudo-spin degree of freedom\cite{sasaki-pseudospin} has also been experimentally tested\cite{georgi2017} and values of a thousand of teslas were found.
In graphene, most of the studies of dynamical deformations\cite{anomalous,moving00,moving01}, have focused on the possibility of generate currents\cite{current01,current02,current03,current04} or topological properties\cite{topo-naumis01,topo-naumis02,topo-naumis03,topo-iade01,topo-iade02} via coherent lattice deformations\cite{topological01} and have considered distortions that depend on time but remain fixed on space\cite{tstatic01,tstatic02}. Instead, here  we consider a traveling deformation\cite{moving00,moving01,moving02,moving03,moving04} and study its effect on the electrons using the Dirac-like equation. The traveling deformation to be considered is an out-of-plane deformation in a tense membrane of graphene. Notice that the physical situation considered here is not to drive graphene in one edge. Instead, graphene can be suspended\cite{electromechanical} with an uniform in-plane strain or with clamped boundaries  such that a displacement of the membrane necessarily imply changing the distance between atoms (strain). Under these circumstances, out-of-plane strain can be induced by an atomic force microscope\cite{nemes}, the electric field from a back-gate electrode\cite{anomalous,Klimov} and/or the tip of a STM probe\cite{georgi2017}. In such circunstances, Klimov et. al found that the deformation can be described as if atoms were moved vertically from their original positions without horizontal shifts\cite{Klimov}, as observed in MD simulations \cite{Monteverde2015}. In our case, the deformations are made time dependent, as for example, by periodically lifting the back-gate electric field. Finally, when graphene lays over a substrate, graphene is usually rippled\cite{quantumwires,Monteverde2015}, and for a given temperature  or for acoustic waves traveling in the substrate, one can produce a deformation as the one considered here.
\begin{figure}[!htbp]
\begin{center}
\includegraphics[scale=0.25]{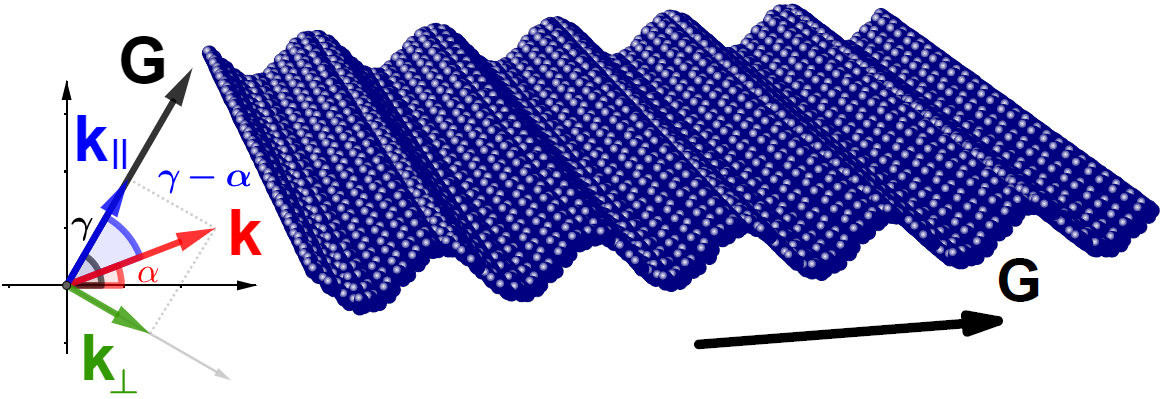}
\end{center}
\caption {(Color online) Diagram of the physical system.
\label{fig01}}
\end{figure}

%
The Hamiltonian for non-interacting electrons in rippled graphene is given by~\cite{sasaki2008pseudospin},
\begin{equation}
H=v_{F}^{0}\boldsymbol{\sigma}_{\eta}\cdot \left(\hat{\boldsymbol{p}}-\eta\boldsymbol{A}(\boldsymbol{r},t\right)) +V(\boldsymbol{r},t),
\end{equation}
such that the dynamical equation is,
\begin{equation}\label{Dirac}
i\hbar\dfrac{\partial}{\partial t}\Psi_{\eta}(\boldsymbol{r},t)  = \left[ v_{F}^{0}\boldsymbol{\sigma}_{\eta}\cdot \left(\hat{\boldsymbol{p}}-\eta\boldsymbol{A}(\boldsymbol{r},t\right) +V(\boldsymbol{r},t) \right] \Psi_{\eta}(\boldsymbol{r},t),
\end{equation}
the subindex $\eta=\pm$ labels the $K$ and $K'$ Dirac points, $v_{F}^{0}$ is the Fermi velocity, $\hat{\boldsymbol{p}}=\left( \hat{p}_{x},\hat{p}_{y} \right)$ is the momentum operator of the charge carriers,   $\boldsymbol{\sigma}_{\eta}=\left( \eta\sigma_{x},\sigma_{y} \right)$ is the vector of Pauli matrices, and $\boldsymbol{A}$ and $V$ are the pseudo-vector and the pseudo-scalar potentials, given by~\cite{sasaki2008pseudospin,neto2009electronic}
 \begin{equation}\label{ScalarField}
  V(\boldsymbol{r},t) =g \left( \varepsilon_{xx}+\varepsilon_{yy} \right), 
  \end{equation}
 \begin{equation}\label{GaugeField}
  \boldsymbol{A }(\boldsymbol{r},t) =\left( A_{x}, A_{y} \right)= \dfrac{\hbar\beta }{2 a_{cc}} \left( \varepsilon_{xx}- 
  \varepsilon_{yy},-2\varepsilon_{xy} 
  \right) .
  \end{equation}
The parameter $g$ ranges from 0 to 20 eV~\cite{vozmediano2010gauge,suzuura2002phonons},   $a_{cc}=1.42 \text{ \AA}$ is the interatomic distance for unstrained pristine graphene, the  dimensionless constant coefficient $\beta\simeq 3.0$ tunes the effect of strain on the hopping parameter. We have neglected the effects from bending\cite{vozmediano2010gauge}, although they can also be described by a gauge field \cite{Amorim}  their effect are about two orders of magnitude smaller. Here we will consider an out-of-plane displacement \(h\), and a in-plane displacement $\boldsymbol{u}$. The strain tensor $\varepsilon_{\mu\nu}$ is given by~\cite{landau-elasticity},
  \begin{equation}\label{strainTensor2}
  \varepsilon_{\mu\nu}=\dfrac{1}{2} \left(\partial_{\mu} h \partial_{\nu} h \right) +\frac{1}{2}\left(\partial_{\mu}\boldsymbol{u}_{\nu}+ \partial_{\nu}\boldsymbol{u}_{\mu}\right).
  \end{equation}

We are considering the most simple propagating deformation, made from a time dependent out-of plane deformation,
\begin{equation}\label{hstrain}
h(\boldsymbol{r},t)=h_{0}\cos{(\boldsymbol{G}\cdot\boldsymbol{r}-\Omega t)},
\end{equation}
where $\boldsymbol{G}$ is the wave-vector of the strain, with components $(G_1,G_2)$. The time-frequency of the strain is given by $\Omega$, which is
related with $G=|\boldsymbol{G}|$ as $\Omega/G=v_s$, where $v_s$ is the strain-propagation velocity, given by the speed of flexural modes. Then we add the possibility of an in-plane strain,
\begin{equation}\label{ustrain}
\boldsymbol{u}(\boldsymbol{r},t)=\boldsymbol{\epsilon}^{0}\cdot\boldsymbol{r}+\boldsymbol{u}^{c}\cos{(\boldsymbol{G}\cdot\boldsymbol{r}-\Omega t))},
\end{equation}
where the first term represents a externally applied uniform strain field, used to avoid excessive bending, plus a second term representing a coupled in-plane propagating strain. Here $\epsilon^{0}$ denotes a space-independent strain tensor, and $\boldsymbol{u}^{c}$ is a constant vector  \cite{naumis-review}. Typical values of these parameters \cite{Monteverde2015,Bai2014} are $h_0 \approx 0.5 nm$ to $h_0 \approx 1 nm$ for small in-plane compressive strains of  $\approx$ 0.6\%, and $G=2\pi/\lambda$, where $\lambda$ ranges from $0.7$ nm to $100$ nm. Notice the huge value of $h_0$ when compared with the in-plane strain; this is due to the strong resistance to compression of C-C bonds \cite{Monteverde2015}.

The fields given by Eq. (\ref{hstrain}) and by Eq. (\ref{ustrain}) are introduced into Eq. (\ref{strainTensor2}) to obtain the deformation tensor. The resulting tensor is thus used in Eq. (\ref{ScalarField}) and Eq. (\ref{GaugeField})  to find the pseudo-vector and pseudo-scalar potentials. Consider first the out-of-plane contribution to the pseudopotentials,
\begin{equation}\label{hScalarField}
  V(\boldsymbol{r},t)=  \bar{g} \sin^2 \phi,
\end{equation}
 \begin{equation}\label{hGaugeField}
  \boldsymbol{A}(\boldsymbol{r},t)=\bar{\beta}
  \left( G_{1}^{2}-G_{2}^{2},-2G_{1}G_{2}\right)\sin^2 \phi.
  \end{equation}
Here we have defined $\bar{g}=(g/2)(G h_0)^2$ and $\bar{\beta}=\hbar \beta h_0^2/(4a_{cc}) $. To simplify notation, we also defined an important variable, the phase of the wave $\phi=\boldsymbol{G}\cdot\boldsymbol{r}-\Omega t$. 

 The in-plane contribution to the pseudopotentials can be calculated in a similar way. However, it is very well known that the constant in-plane strain leads to a renormalized-direction-dependent Fermi velocity\cite{naumis-review}. Assuming that the axis coincide with the principal directions of the strain, the Fermi velocity is now a diagonal tensor with components $(\boldsymbol{v}_F)_{\mu\mu}=[1+(1-\beta)\epsilon^{0}_{\mu \mu}]v_F^{0}$. To keep the equations simple, and since we can chose externally the strain, we suppose that the field is such that the Fermi velocity is the same in the $x$ and $y$ directions. Then we can use Eq. (\ref{Dirac}) by making the renormalization $v_F^{0} \rightarrow v_F= [1+(1-\beta)\epsilon^{0}_{xx}]v_F^{0}$. Let us now consider the coupled in-plane strain potentials. This results in a  new contribution $V_c(\boldsymbol{r},t)=  \bar{g}_c \sin\phi$ with $\bar{g}_c=g\boldsymbol{u}^{c}\cdot \boldsymbol{G}$, and
\begin{equation}
\boldsymbol{A}_c(\boldsymbol{r},t)=\bar{\beta}_{c}(G^{2}\cos\theta^{c},-2GG_{2}u_{x}^{c}/u^{c})\sin\phi.
\end{equation}
Here  $\bar{\beta}_c=\hbar \beta u^{c}/(2a_{cc}G)$ and $\theta^{c}$ is the angle between $\boldsymbol{u}^{c}$ and $\boldsymbol{G}$. Although, $h_0$ enters quadratically in the potentials while $u^{c}$  linearly, we are considering out-of-plane deformations, where $h_0/u^{c}\gg 1$ such that $\bar{g}\gg \bar{g}_c$ and $\bar{\beta}\gg\bar{\beta}_c$. Thus the in-plane coupled part can be neglected.  Eventually, it can be included following the same procedure considered here.

Finally, since $ (G_{1} \pm i\eta G_{2})^2=G_{1}^{2}-G_{2}^{2}\pm 2i\eta G_1G_2$, the Dirac equation  coupled with the propagating deformation is,
\begin{equation}\label{paired-one}
\begin{split}
\dfrac{\partial}{\partial t}\psi_{A}=&\dfrac{\bar{g}}{i\hbar}\sin^2\phi\psi_A-v_{F}\bigg[ \eta\partial_{x} \mp i\partial_{y}\\
& + \dfrac{\bar{\beta}}{i\hbar}\sin^2(\phi)
(G_{1}\pm i\eta G_{2})^2 \bigg]\psi_{B}.
\end{split}
\end{equation}
The lower signs are used to obtain a second equation, made by replacing $A \rightarrow B$ and $B \rightarrow A$ in Eq. (\ref{paired-one}). $\psi_{A}$ and $\psi_{B}$ are the two components of the spinor $\Psi_{\eta}(\boldsymbol{r},t)$. They represent the wavefunction components of the electron in each of the graphene's triangular sublattices $A$ and $B$. 
The most important step in the solution of this problem is to propose a solution of the form~\cite{landau1959course},
\begin{equation}\label{wavefunction}
\boldsymbol{\psi}_{\rho}=\exp{\left[i\boldsymbol{k}\cdot\boldsymbol{r}-i\dfrac{Et}{\hbar}\right]}\Phi_{\rho} (\phi),
\end{equation}
where $\Phi_{\rho} (\phi)$ is a function to be determined for $\rho=A,B$, and $E$ is an energy related to the momentum $\boldsymbol{p}=\hbar \boldsymbol{k}$ by $E=v_{F}\hbar |\boldsymbol{k}|$. This ansatz is equivalent to consider the problem in the space-time frame of the moving wave \cite{landau1960course,moving02}. As detailed in the supplementary section, the system of differential equations can be further rewritten in terms of two new functions $\Gamma_A(\phi)$ and $\Gamma_B(\phi)$, defined by,
\begin{equation}
\Gamma_{\rho} =e^{i[\eta \gamma/2+\pi/4+\tilde{k}_{||}\phi-\eta \tilde{A_0}\cos(3\gamma)(\phi-\sin\phi\cos\phi)/2 ] }\Phi_{\rho},
\end{equation}
and obtain,
\begin{equation}\label{AuxA}
\dfrac{d\Gamma_A(\phi)}{d\phi}= D(\phi)\Gamma_A(\phi)+C(\phi)\Gamma_B(\phi), 
\end{equation}
\begin{equation}\label{AuxB}
\dfrac{d\Gamma_{B}(\phi)}{d\phi}= -C(\phi)\Gamma_A(\phi) -D(\phi) \Gamma_B(\phi) ,
\end{equation}
where $C(\phi)$ and $D(\phi)$ are defined as,
\begin{equation} \label{Eq:DefinitionC}
C(\phi)= \eta(|\tilde{\textbf{k}}| -\tilde{g}\sin^2\phi),
\nonumber
\end{equation}
\begin{equation} 
D(\phi)= \left[\tilde{A_0} \sin(3\gamma)\sin^2\phi-\eta\tilde{k}_{\perp}\right], 
\nonumber
\end{equation}
and $\tilde{A_{0}}=\bar{\beta}G/\hbar$ and $\tilde{g}=\bar{g}/v_F \hbar  G$.
Here the vector $\tilde{\textbf{k}}=(\tilde{k}_x,\tilde{k}_y)$, defined as  $\tilde{\textbf{k}}=\textbf{k}/G$,  was projected into the parallel and  perpendicular directions of the propagating corrugation (see supplementary material),
\begin{equation}
\tilde{k}_{||}=\tilde{k}\cos(\gamma-\alpha),\tilde{k}_{\perp}=\tilde{k}\sin(\gamma-\alpha),
\nonumber
\end{equation}
where $\gamma=\tan^{-1}\left(G_2/G_1\right)$ is the angle between the $x$ axis (graphene's zigzag direction) and the propagating direction of the flexural mode. The angle $\alpha$ is the direction of the momentum given by $\alpha=\tan^{-1}\left(\eta k_y/k_x\right)$.

As explained in the supplementary material,  by elimination of $\Gamma_{B}(\phi)$, Eqns. (\ref{AuxA}) and (\ref{AuxB}) can be written as a single second order ordinary differential equation. In the resulting equation, the first derivative of $\Gamma_{A}(\phi)$ can be further eliminated by using the ansatz,
\begin{equation}
\Gamma_A(\phi)=\frac{Z(\phi)}{\sqrt{C(\phi)}},
\end{equation}
where $Z(\phi)$ follows a Hill's equation,
\begin{equation}\label{Hill}
\frac{d^2 Z(\phi)}{d^2\phi}+F(\phi)Z(\phi)=0,
\end{equation}
with $F(\phi)$ defined as,
\begin{equation}\label{Fofphi}
\begin{split}
F(\phi)&=-\left[D'(\phi)-\frac{C'(\phi)}{C(\phi)}D(\phi)-C^2(\phi) \right. \\
& +D^2(\phi)\bigg]+ \frac{C''(\phi)}{2C(\phi)}-\frac{3}{4}\left(\frac{C'(\phi)}{C(\phi)}\right)^2.
\end{split}
\end{equation}

The resulting Hill equation is difficult to be  solved analytically for all cases. Yet there are important solvable limiting cases. Let us first consider a pseudo-magnetic field without a pseudo-scalar field. In this case, we have $\tilde{g}=0$ and $C(\phi)=\eta |\tilde{\boldsymbol{k}}|$. Therefore, from  Eq. (\ref{Hill}) we obtain, 
\begin{equation}\label{Eq:magnetic}
-\frac{d^2\Gamma_A(\phi)}{d\phi^2}  +\left[\frac{d}{d\phi}D(\phi)-|\tilde{\boldsymbol{k}}|^2  +D^2(\phi)\right]\Gamma_A(\phi)=0. 
\end{equation}
Since the flexural mode amplitude is small, in Eq. (\ref{Eq:magnetic}) we can neglect the quadratic term in $\tilde{A_0}$. Eq. (\ref{Eq:magnetic}) is thus transformed into the following Mathieu equation,
\begin{equation}
\frac{d^2\Gamma_A(\zeta_{\pm})}{d\zeta^2_{\pm}}+\left[a_{\pm}-2q cos (2\zeta_{\pm}) \right]\Gamma_A(\zeta_{\pm})=0, 
\end{equation}
by using a change of variables from $\phi$ to $\zeta_{+}$ and $\zeta_{-}$. For the valley $\eta=1$, the variable $\zeta_{+}$ must be used,
\begin{equation}
\zeta_{+}=\phi - \phi_0  ,  
\end{equation}
while for the valley $\eta=-1$, $\zeta_{-}$ must be used
\begin{equation}
\zeta_{-}=\phi + \phi_0,
\end{equation}
The parameters  $\phi_0$ and $q$ are defined as,
\begin{equation}\label{Eqqvectorial}
\tan  (2\phi_0)=\frac{1}{\tilde{k}_{\perp}}, \ q=\frac{\tilde{A_0}}{2}\sin(3\gamma)\sqrt{1+\tilde{k}_{\perp}^2} 
\end{equation}
while $a_{\pm}$ is defined as,
\begin{equation}\label{Eqavectorial}
a_{\pm}=\tilde{k}_{||}^2+\eta \tilde{k}_{\perp}\tilde{A_0}sin(3\gamma), 
\end{equation}

\begin{figure}[!htbp]
\begin{center}
\includegraphics[scale=0.45]{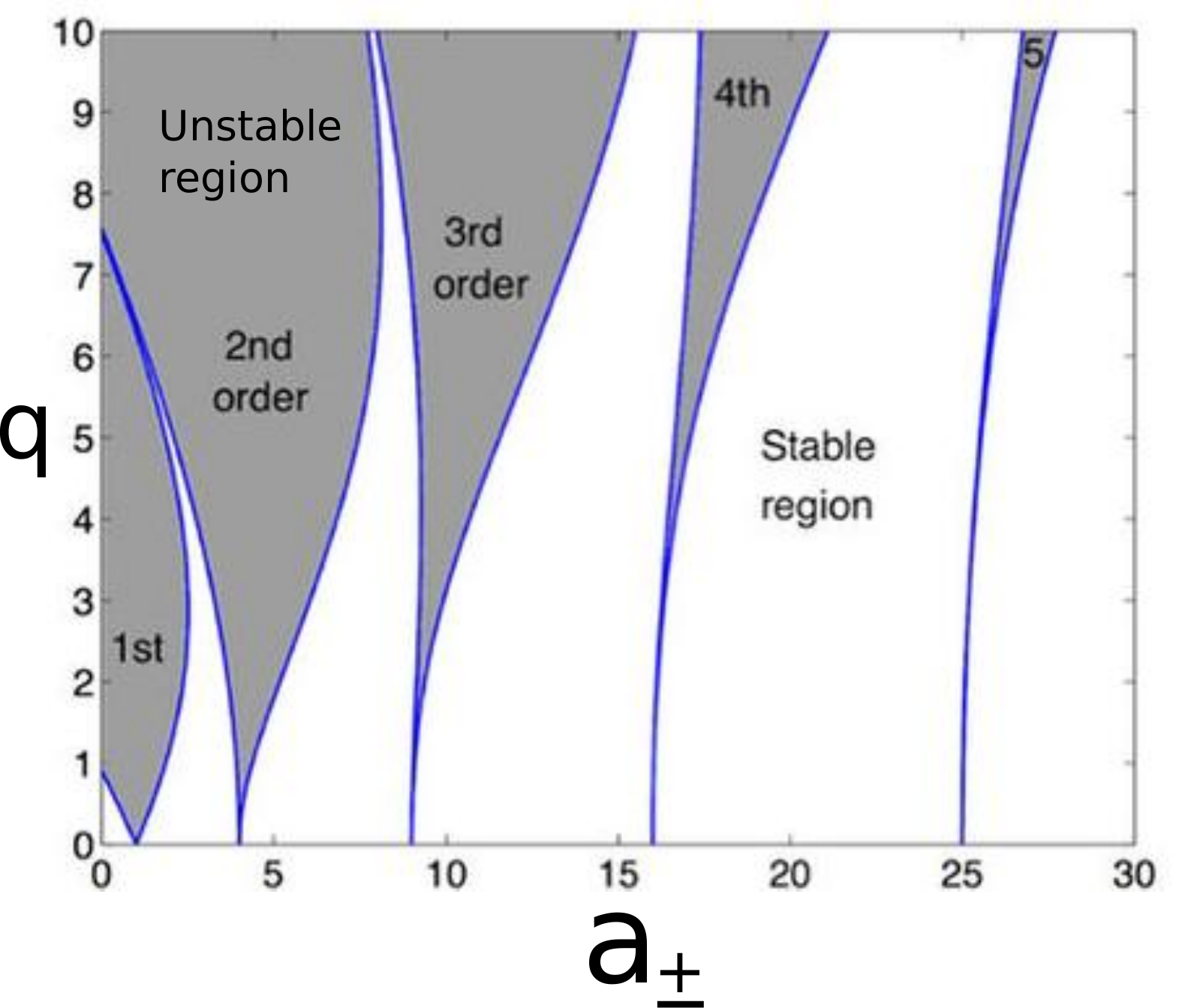}
\end{center}
\caption {(Color online) Spectrum of the Mathieu equation showing the allowed and forbidden (shaded area) regions. This spectrum is valid for the pure pseudovectorial case (see Eqns. (\ref{Eqqvectorial}) and (\ref{Eqavectorial}) ) or for the pseudovectorial and pseudoscalar case with $|\boldsymbol{\tilde{k}}|>>\tilde{A}_{0}$ and$|\boldsymbol{\tilde{k}}|>>\tilde{g}$ (see Eqns. (\ref{Eqaboth}) and (\ref{Eqqboth})). The order of the gaps are indicated.
\label{FigMathieu}}.
\end{figure}

The are some important remarks concerning this equation. It describes a well known classical problem: a parametric pendulum. In such pendulum, the length  is changed periodically resulting in a pattern of resonances. Fig. \ref{FigMathieu} present the allowed regions for stable solutions of the Mathieu equation, which for this problem indicates that the spectrum is made of bands and gaps. To understand the gap opening, suppose that $\tilde{k}_{\perp}=0$. As seen in Fig.~\ref{FigMathieu}, for $q\ll 1$ gaps are open whenever $a_{\pm}=\tilde{k}_{||}^2\approx n^2\ $ with $n=1,2,3,...$ resulting in
$|\boldsymbol{k}| \approx n G$, which is a diffraction condition due to the ripple wave-periodicity. Translated into energy, this condition is  $E\approx nv_{F} \hbar G$, which is the energy of an electron with  momentum $G$.  In graphene, $\hbar v_F\approx 0.65775$ eV.nm, and $E\approx (4.13/\lambda)$ eV.nm if $\lambda$ is given in nanometers \cite{naumis-review}. For example, $E\approx 0.0413$ eV for a ripple with $\lambda=100$ nm. As $h_{0}$ grows, $E$ decreases from this value. For $q=1$, a gap opens at the Dirac point. On the other hand, if $|\boldsymbol{\tilde{k}}| \approx \tilde{k}_{\perp}\gg 1$, then $a_{\pm}\approx q$ and the gaps become much wider, meaning that in general, the propagation is preferentially in a direction parallel to the ripple. For a fixed value of $a_{\pm}$ and $q$, the general solution is a linear combination of the Mathieu  cosine $C(a_{\pm},q,\zeta)$ and Mathieu sine $S(a_{\pm},q,\zeta)$ functions. In this particular case, the solution must reduce to the free-particle wavefunction for $\tilde{A_0}=0$. By taking into account that $C(a_{\pm},0,\zeta)=\cos(\sqrt{a_{\pm}}\zeta)$ and $S(a_{\pm},0,\zeta)=\sin(\sqrt{a_{\pm}}\zeta)$, the solution 
can be written as,
\begin{equation}
\Gamma(\zeta_{\pm})=\left[C(a_{\pm},q,\zeta_{\pm})+iS(a_{\pm},q,\zeta_{\pm})\right]\left( {\begin{array}{cc}
   1\\
   s \exp(i\alpha)\\
  \end{array} } \right).
\end{equation}
where $s=\pm 1$, where the minus is used for the conduction and valence bands respectively. The second observation is that solutions on each valley are out of phase by a factor $2\phi_0$. This is an suggests a kind of quantum pump that deserves further investigation.

Let us now analyze the presence of both pseudo scalar and pseudo vectorial  fields. The expression for $F(\phi)$ is in general too complicated to extract the physical behavior behind it. Instead consider first the limit $\tilde{|\boldsymbol{k}|}>>\tilde{g}$ and $\tilde{|\boldsymbol{k}|}>>\tilde{A}_0$. Under such assumption $F(\phi)$ is given by,
\begin{equation}
F(\phi)\approx \frac{C'(\phi)}{C(\phi)}D(\phi)+C^2(\phi)-D'(\phi)-D^2(\phi)-\frac{C''(\phi)}{2C(\phi)},
\end{equation}
from where it follows that,
\begin{equation}
F(\phi)\approx a_{\pm}+L(\tilde{k})cos(2\phi)+M(\tilde{k})sin(2\phi),
\end{equation}
with,
\begin{equation}\label{Eqaboth}
a_{\pm}=\tilde{k}_{||}^2+\eta \tilde{k}_{\perp}\tilde{A_{0}}\sin(3\gamma)-\tilde{g}|k|,
\end{equation}
\begin{equation}
L(\tilde{k})=-\eta \tilde{k}_{\perp}  \tilde{A_0}\sin(3\gamma)+\tilde{g}
\left( | \tilde{\boldsymbol{k}}|-\frac{1}{| \tilde{k}|}\right),
\end{equation}
\begin{equation}
M(\tilde{k})=-\tilde{A_0}\sin(3\gamma)+\eta \tilde{g} \sin(\gamma-\alpha),
\end{equation}
By setting,
\begin{equation}
L(\tilde{k})\cos(2\phi)+M(\tilde{k})\sin(2\phi)=-2q\cos(\phi+ \phi_0),
\end{equation}
where $q$ and $\phi_0$ are given by, 
\begin{equation}\label{Eqqboth}
q=\frac{\sqrt{L^2(\tilde{k})+M^2(\tilde{k})}}{2},
\end{equation}
and 
\begin{equation}
\tan(\phi_0)=-\frac{M(\tilde{k})}{L(\tilde{k})}.
\end{equation}
Defining a new variable $\zeta=\phi+\phi_0$ which takes into account the phase shift,  again we obtain a Matthieu equation for $Z(\zeta)$,
\begin{equation}
\frac{d^2Z(\zeta)}{d\zeta^22}+(a_{\pm}-2q\cos(2\zeta))Z(\zeta)=0.
\end{equation}
Notice that for $\tilde{g}=0$, we recover exactly the case without the scalar field, while for $\tilde{g}\neq 0$, $a_{\pm}$ and $q$ are modified to include an
energy correction due to the pseudoscalar field.  Once the solution for $Z(\phi)$ is found by using the Matthieu functions, we obtain that, 
\begin{equation}
\Gamma_A(\phi)\approx \left( 1+\frac{\tilde{g}}{\tilde{k}}\sin^2(\phi)\right)\frac{Z(\phi)}{\tilde{k}}.
\end{equation}

When $| \tilde{\boldsymbol{k}}|\le \tilde{g}$, the previous approximations brake down. Thus a different approach is needed. For simplicity, consider the case $D(\phi)=0$. Instead of using Eq. (\ref{Fofphi}), it is easier to start from Eqns. (\ref{AuxA}) and (\ref{AuxB}), 
\begin{equation}
\Gamma_{A}''(\phi)+\frac{1}{C(\phi)}\frac{dC(\phi)}{\phi}\Gamma_{A}'(\phi)+C(\phi))^2\Gamma_{A}(\phi)=0,
\end{equation}
resulting in, 
\begin{equation}
\Gamma_{A}''(\phi)+\frac{q \sin(2\phi)}{\epsilon+q\cos(2\phi)}\Gamma_{A}'(\phi)+(\epsilon+q\cos(2\phi))^2\Gamma_{A}(\phi)=0,
\end{equation}
with $\epsilon=\tilde{k}-\tilde{g}/2$ and $q=\tilde{g}/2$.
By making the substitution $\Gamma_A(\phi)=u(q\sin(2\phi)+2\epsilon \phi)$, it follows that $u$ satisfies the harmonic oscillator equation, from where the solution is given by,
\begin{equation}\label{Eq:ExactPseudo}
\begin{split}
\Gamma_A(\phi)&=C_1\cos\left[\frac{\tilde{g}}{4}\sin(2\phi)+(\tilde{k}-\tilde{g}/4)2\phi\right]\\
&+C_2\sin \left[\frac{\tilde{g}}{4}\sin(2\phi)+(\tilde{k}-\tilde{g}/4)2\phi\right],
\end{split}
\end{equation}
where $C_1$ and $C_2$ are constants determined from the initial conditions. 

\begin{figure}[!htbp]
\begin{center}
\includegraphics[scale=0.34]{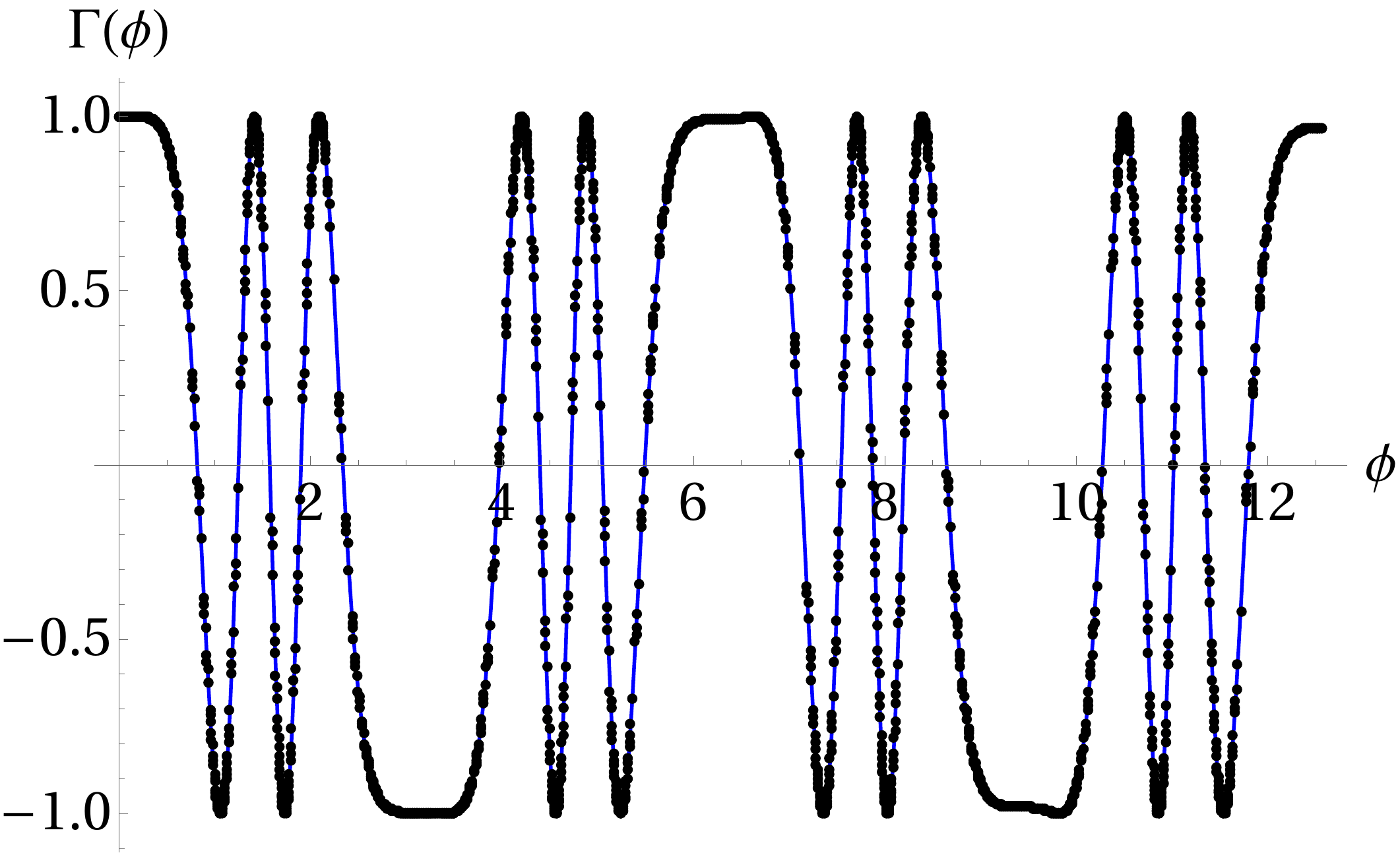}
\end{center}
\caption {(Color online)  $\Gamma_A(\phi)$ for a pure pseudoscalar field  obtained from the exact solution (continuous   blue curve) given by Eq. (\ref{Eq:ExactPseudo}), compared with a numerical solution obtained using a Runge-Kutta algorithm (points), for $\tilde{E}=\tilde{k}=0$. There is a perfect matching between both.  The corresponding parameters are $\tilde{A}_0=0,\tilde{k}_{\perp}=0$, for $\tilde{k}=0$, and  for $\tilde{g}=\bar{g}/ v_F \hbar G=10$}. 
\label{FigExactVsNumeric2}.
\end{figure}

Figure \ref{FigExactVsNumeric2} shows a comparison between the analytical solution Eq. (\ref{Eq:ExactPseudo}) and a numerical solution of Eqns. (\ref{AuxA}) and (\ref{AuxB}), obtained using a Runge-Kutta algorithm, showing a  perfect match.
Figure \ref{FigExactVsNumeric2} is the solution for $\tilde{E}=\tilde{k}=0$. In this plot $\tilde{g}=10$, corresponding to $h_0 \approx 2$ \ nm and $\lambda \approx 39$ \ nm. This value has been chosen because it is within the limits of the physical system and the Dirac approximation. To see this, consider that $\bar{g}=(g/2)(Gh_0)^2$, and since $G=2\pi/\lambda$, we obtain $\bar{g}=(2\pi ^2g)(h_0/\lambda)^2$. Then, $\tilde{g}=(2\pi ^2g)(h_0/\lambda)^2/[ v_F \hbar 2 \pi/ \lambda]$ . The biggest estimation of $g$ is $20$ eV, resulting in $\bar{g}\approx 395(h_0/\lambda)^2\text{eV}$. Finally, $\tilde{g}\approx 95 (h_0^2/\lambda)\text{nm}^{-1}$. Assuming $h_0 \approx 1 \ nm$, and the minimal $\lambda \approx 1 \ nm$, we have  $\tilde{g} \approx 95$. From there, $\tilde{g}$ goes to zero as $\lambda \rightarrow \infty$ and $h_0 \rightarrow 0$. However, the Dirac approximation imposes $h_0/\lambda<0.053$, resulting in the limit 
$\tilde{g} < 25$ to our approach.

In both solutions, we observe regions of oscillations with a frequency determined by $\tilde{g}$ and $\tilde{k}$. Therein, the electrons are acelerated, while for certain phases, the solutions present a maximum with a wider width. It is important to remark that our solution implies the generation of high-harmonics in response to the field. To show this, assume that $C_1=1$ and $C_2=0$ in Eq. (\ref{Eq:ExactPseudo}). Defining $\tilde{\omega}=2(\tilde{k}-\tilde{g}/4)$, using a trigonometric identity and by expanding the composition of trigonometric functions in terms of the Bessel functions $J_p(\tilde{g}/4)$ we obtain,

\begin{equation}
\begin{split}
\Gamma_A(\phi)&=\left(J_{0}(\tilde{g}/4)+\sum_{p=1} J_{2p}(\tilde{g}/4)cos(4p\phi)\right)\cos[\tilde{\omega}\phi]\\
&-\left(\sum_{p=0} J_{2p+1}(\tilde{g}/4)cos((2p+1)2\phi)\right)\sin[\tilde{\omega}\phi].
\end{split}
\end{equation}

Thus, the electron dynamic response contains a rich structure of harmonics. Observe also that the zeros of the Bessel functions allows to tune the pseudoscalar field in such a way that certain harmonics can be eliminated.

\begin{figure}[!htbp]
\begin{center}
\includegraphics[scale=0.30]{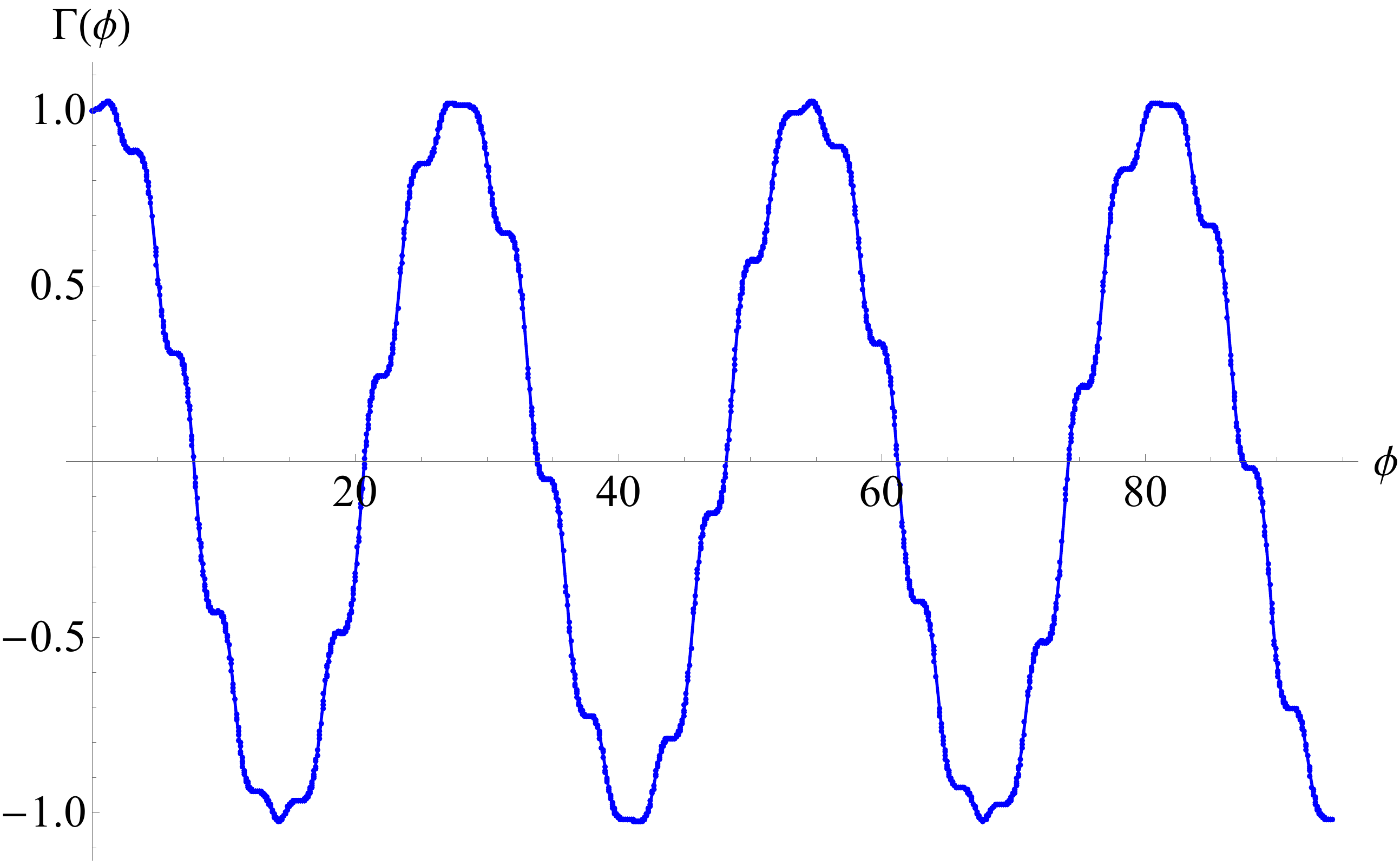}
\end{center}
\caption {(Color online) Numerical solution obtained using a Runge-Kutta algorithm (blue lines) in the case of an applied pseudovectorial and pseudoscalar fields. The corresponding parameters are $\tilde{A}_0=0.1,\tilde{k}_{\perp}=0.01$, for $\tilde{E}=\tilde{k}_{\perp}$, with $\tilde{g}=\bar{g}/v_F \hbar  G=0.5$. 
\label{FigBothFields}}
\end{figure}

We end up by showing in Fig. \ref{FigBothFields} a numerical solution of the case in which both pseudoscalar and pseudovectorial fields are present. The mains differences are the modification of the oscillating frequency and that the regions of nearly constant amplitude present more structure, instead of a maximum.

In conclusion, we have solved the problem of electrons in graphene under a propagating ripple. This was done by solving a Dirac equation which includes the corresponding propagating pseudoscalar and pseudovectorial fields. In the presence of a ripple with only a pseudovectorial field, the equation can be transformed into an ordinary second-order partial equation, i.e., into the Matthieu equation. Such system presents a spectrum made from gaps and bands. Gaps are basically determined from electron diffraction by ripple waves. Also, electron propagation happens mainly in the ripple-propagation-direction. For the presence of both a pseudovectorial and pseudoscalar field, we also found a Matthieu equations with a spectrum of bands and gaps but with different parameters. However, when the energy of the electron is less than the energy associated to the pseudoscalar field, a different kind of equation is obtained. For the pure pseudoscalar field, no gaps are observed. In all cases, the solutions contain high-harmonics with respect to the fundamental frequency of the driving field. 

\section*{Acknowledgements}
R.C.B acknowledges useful discussions with M. Asmar and I. Vekhter. This work was supported by project UNAM-DGAPA-PAPIIT-IN102717 and Movilidad Acad\'emica 2017 (UABC).

\appendix

\onecolumngrid

\section{Obtention of Eqs. (14) and (15)}

In this section of the supplementary material, we present the detailed steps needed to obtain Eqns. (14) and (15) from Eq. (12) in the main text. First we calculate all the required partial derivatives of the wavefunction ansatz given by Eq. (12),
\begin{equation}\label{timederivative}
\dfrac{\partial}{\partial t}\psi_{\rho}=e^{i\boldsymbol{k}\cdot\boldsymbol{r}}e^{-i\frac{Et}{\hbar}}\left[ \dfrac{-iE}{\hbar}  \Phi_{\rho}-\Omega \dfrac{d\Phi_\rho}{d\phi} \right]
\end{equation}
\begin{equation}\label{xderivative}
\dfrac{\partial}{\partial x}\psi_{\rho}=e^{i\boldsymbol{k}\cdot\boldsymbol{r}}e^{-i\frac{Et}{\hbar}}\left[ ik_{x} \Phi_{\rho}+ G_{1}\dfrac{d\Phi_{\rho}}{d\phi} \right]
\end{equation}
\begin{equation}\label{yderivative}
\dfrac{\partial}{\partial y}\psi_{\rho}=e^{i\boldsymbol{k}\cdot\boldsymbol{r}}e^{-i\frac{Et}{\hbar}}\left[ ik_{y} \Phi_{\rho}+ G_{2}\dfrac{d\Phi_{\rho}}{d\phi} \right]
\end{equation}
After substituting these partials in Eq. (11), we get two equations for 
$\Phi_A$ and $\Phi_B$,
\begin{equation}\label{coupled-A}
\begin{split}
 -\dfrac{iE}{\hbar}  \Phi_A &-\Omega \dfrac{d\Phi_A}{d\phi} =\dfrac{\bar{g}}{i\hbar}\sin^2\phi\Phi_A \\
  &-v_{F} \eta\left( ik_{x} \Phi_{B}+ G_{1}\dfrac{d\Phi_{B}}{d\phi} \right)\\
  &-iv_{F}\left( ik_{y} \Phi_{B}+ G_{2}\dfrac{d\Phi_{B}}{d\phi} \right)\\ 
  &+v_{F}\dfrac{\bar{\beta}}{i\hbar} \sin^2\phi (G_{1}+ i\eta G_{2})^2\Phi_B  ,
\end{split}
\end{equation}
\begin{equation}\label{coupled-B}
\begin{split}
 -\dfrac{iE}{\hbar}  \Phi_B &-\Omega \dfrac{d\Phi_B}{d\phi} =\dfrac{\bar{g}}{i\hbar}\sin^2\phi\Phi_B \\
  &-v_{F}  \eta\left( ik_{x} \Phi_{A}+ G_{1}\dfrac{d\Phi_{A}}{d\phi} \right)\\
  &+iv_{F}\left( ik_{y} \Phi_{A}+ G_{2}\dfrac{d\Phi_{A}}{d\phi} \right) \\+ &v_{F}\dfrac{\bar{\beta}}{i\hbar} \sin^2\phi (G_{1}- i\eta G_{2})^2\Phi_A  ,
\end{split}
\end{equation}
Let $\gamma=\tan^-1\left(G_2/G_1\right)$ such that $G_1=G\cos\gamma$ and $G_2=G\sin\gamma$ with $G=\sqrt{G_1^2+G_2^2}$. With these definitions we have 
\begin{equation}
G_1\pm i\eta G_2=G\left(\cos\gamma\pm i\eta \sin\gamma\right)=Ge^{\pm i\eta\gamma}
\end{equation}
such that
\begin{equation}
\left(G_1\pm i \eta G_2 \right)^2=G^2e^{\pm 2 i\eta\gamma}.
\end{equation}
Using the previous expression, we can rewrite Eq. (\ref{coupled-A}) and (\ref{coupled-B}) as follows,

\begin{equation}\label{Approximated-new-B}
\begin{split}
 -i\tilde{E}\Phi_A & =-i\tilde{g}\sin^2\phi\Phi_A \\
  &-i\eta \tilde{k}_{x}\Phi_{B} -\tilde{k}_y\Phi_{B} -\eta e^{-i\eta\gamma} \dfrac{d\Phi_{B}}{d\phi} +i A_0 e^{+i2\eta\gamma} \sin^2\phi  \Phi_B  ,
\end{split}
\end{equation}
\begin{equation}\label{Approximated-new-A}
\begin{split}
 -i\tilde{E}\Phi_B &=-i\tilde{g} \sin^2\phi\Phi_B \\
  &-i\eta \tilde{k}_x\Phi_{A} +\tilde{k}_y\Phi_{A} -\eta e^{i\eta\gamma} \dfrac{d\Phi_{A}}{d\phi} +i A_0 e^{-i2\eta\gamma}\sin^2\phi \Phi_A ,
\end{split}
\end{equation}
where we have defined $\tilde{E}=\dfrac{E}{\hbar v_F G}$, $\tilde{g}=\dfrac{\bar{g}}{\hbar v_F G}=$, $\tilde{k_x}=\dfrac{ k_x}{G} $, $\tilde{k_y}=\dfrac{ k_y}{G} $ and 
$A_0=\dfrac{\bar{\beta}G}{\hbar }$. Also, we have used that since $\dfrac{\Omega}{G}=V_{s}$, we have $\dfrac{\Omega}{v_F G}=\dfrac{v_s}{v_F}\ll 1$ and we can neglect the second term on each equation,

We can rewrite these equations in the following manner,
\begin{equation}\label{fixed-new-B2}
 \dfrac{d\Phi_{B}}{d\phi}  =i \eta e^{i\eta\gamma}\left( \tilde{E} -\tilde{g}\sin^2\phi \right)\Phi_A 
  -i e^{i\eta\gamma} \left( \tilde{k}_{x} -i\eta\tilde{k}_y -\eta A_0 e^{+i2\eta\gamma} \sin^2\phi\right)  \Phi_B  ,
\end{equation}
\begin{equation}\label{fixed-new-A2}
 \dfrac{d\Phi_{A}}{d\phi}  =i \eta e^{-i\eta\gamma}\left( \tilde{E} -\tilde{g}\sin^2\phi \right)\Phi_B 
  -i e^{-i\eta\gamma} \left( \tilde{k}_{x} +i\eta\tilde{k}_y -\eta A_0 e^{-i2\eta\gamma} \sin^2\phi\right)  \Phi_A  ,
\end{equation}
the terms with the momentum components can be written as
\begin{equation}
\tilde{k}_{x} \pm i\eta\tilde{k}_y=\tilde{k}e^{\pm i\eta\alpha}
\end{equation}
and then 
\begin{equation}\label{fixed-new-B}
 \dfrac{d\Phi_{B}}{d\phi}  =i \eta e^{i\eta\gamma}\left( \tilde{E} -\tilde{g}\sin^2\phi \right)\Phi_A 
  -i e^{i\eta\gamma} \left( \tilde{k}e^{- i\eta\alpha} -\eta A_0 e^{+i2\eta\gamma} \sin^2\phi\right)  \Phi_B  ,
\end{equation}
\begin{equation}\label{fixed-new-A}
 \dfrac{d\Phi_{A}}{d\phi}  =i \eta e^{-i\eta\gamma}\left( \tilde{E} -\tilde{g}\sin^2\phi \right)\Phi_B 
  -i e^{-i\eta\gamma} \left( \tilde{k}e^{ i\eta\alpha} -\eta A_0 e^{-i2\eta\gamma} \sin^2\phi\right)  \Phi_A  ,
\end{equation}
We apply the following transformation,
\begin{equation}\label{toTransformation-Chi}
\begin{split}
 \chi_A &=\exp{(\dfrac{i\eta \gamma}{2})}\exp{(\dfrac{i\pi}{4})}\Phi_A \\
  \chi_B&=-\exp{(\dfrac{-i\eta \gamma}{2})}\exp{(\dfrac{-i\pi}{4})}\Phi_A,
\end{split}
\end{equation}
or
\begin{equation}\label{toTransformation-Chi2}
\begin{split}
 \Phi_A &=\exp{(\dfrac{-i\eta \gamma}{2})}\exp{(\dfrac{-i\pi}{4})}\chi_A \\
  \Phi_B&=-\exp{(\dfrac{i\eta \gamma}{2})}\exp{(\dfrac{i\pi}{4})}\chi_A.
\end{split}
\end{equation}
to Eq. \ref{fixed-new-B} and Eq.\ref{fixed-new-A} to obtain,

\begin{equation}\label{chi-B}
 -\dfrac{d\chi_{B}}{d\phi}  =\eta\left( \tilde{E} -\tilde{g}\sin^2\phi \right)\chi_A + i \tilde{k} e^{i\eta(\gamma-\alpha)}\chi_B-i\eta e^{+i3\eta\gamma} A_0  \sin^2\phi  \chi_B  ,
\end{equation}
\begin{equation}\label{chi-A}
 \dfrac{d\chi_{A}}{d\phi}  = \eta\left( \tilde{E} -\tilde{g}\sin^2\phi \right)\chi_B -i \tilde{k} e^{-i\eta(\gamma-\alpha)}  \chi_A +i\eta e^{-i3\eta\gamma} A_0  \sin^2\phi\chi_A ,
\end{equation}
We expand the exponentials in terms of sines and cosines,
\begin{equation}\label{chi-B-s}
\begin{split}
 -\dfrac{d\chi_{B}}{d\phi}  =&\eta\left( \tilde{E} -\tilde{g}\sin^2\phi \right)\chi_A + i \tilde{k} \cos(\gamma-\alpha)\chi_B \\
 &-\eta \tilde{k} \sin(\gamma-\alpha)\chi_B-i\eta A_0 \cos(3\gamma)\sin^2\phi  \chi_B+ A_0\sin(3\gamma)  \sin^2\phi  \chi_B  ,
\end{split}
\end{equation}
\begin{equation}\label{chi-B-s2}
\begin{split}
 -\dfrac{d\chi_{A}}{d\phi}  =&\eta\left( \tilde{E} -\tilde{g}\sin^2\phi \right)\chi_B - i \tilde{k} \cos(\gamma-\alpha)\chi_A \\
 &-\eta \tilde{k} \sin(\gamma-\alpha)\chi_A+i\eta A_0 \cos(3\gamma)\sin^2\phi  \chi_A+ A_0\sin(3\gamma)  \sin^2\phi  \chi_A  ,
\end{split}
\end{equation}
To eliminate the terms with imaginary coefficients in the previous equations, we propose to use the following transformation,
\begin{equation}
\chi_{\rho}=\exp\left[-i\tilde{k}\cos(\gamma-\alpha)\phi+i\eta A_0\cos(3\gamma)\left(\dfrac{1}{2}\phi-\dfrac{1}{2}\sin\phi\cos\phi \right) \right]\Gamma_{\rho}
\end{equation}
from where we obtain the following set of equations,
\begin{equation}\label{gamma-B}
 -\dfrac{d\Gamma_{B}}{d\phi}  =\eta\left( \tilde{E} -\tilde{g}\sin^2\phi \right)\Gamma_A -\eta \tilde{k} \sin(\gamma-\alpha)\Gamma_B+A_0\sin(3\gamma)\sin^2\phi \Gamma_B  ,
\end{equation}
\begin{equation}\label{gamma-A}
 \dfrac{d\Gamma_{A}}{d\phi}  =\eta\left( \tilde{E} -\tilde{g}\sin^2\phi \right)\Gamma_B -\eta \tilde{k} \sin(\gamma-\alpha)\Gamma_A+A_0\sin(3\gamma)\sin^2\phi \Gamma_A,
\end{equation}
or
\begin{equation}\label{gamma-A-f}
 \dfrac{d\Gamma_{A}}{d\phi}  = \left[A_0\sin(3\gamma)\sin^2\phi-\eta \tilde{k} \sin(\gamma-\alpha) \right]\Gamma_A+ \eta\left( \tilde{E} -\tilde{g}\sin^2\phi \right)\Gamma_B,
\end{equation}
\begin{equation}\label{gamma-B-f}
 \dfrac{d\Gamma_{B}}{d\phi}  = 
 -\eta\left( \tilde{E} -\tilde{g}\sin^2\phi \right)\Gamma_A- \left[A_0\sin(3\gamma)\sin^2\phi -\eta \tilde{k} \sin(\gamma-\alpha) \right] \Gamma_B  ,
\end{equation}

Using the dispersion relation $\tilde{E}=|\tilde{\boldsymbol{k}}|$, the final Eqns.  (14) and (15)  are obtained.

\section{Reduction of the coupled differential equations to an ordinary differential equation}

Let us now study the properties of  Eq. (14) and (15). This system can be written in terms of a matrix $\boldsymbol{J}(\phi)$ acting on a vector, 
\begin{equation}
\frac{d\boldsymbol{\Gamma}(\phi)}{d\phi}=\boldsymbol{J}(\phi)\boldsymbol{\Gamma}
\end{equation}
where $\boldsymbol{\Gamma(\phi)}=(\Gamma_A(\phi),\Gamma_B(\phi))$ and,
\begin{equation}
\boldsymbol{J}(\phi)=
\left[ {\begin{array}{cc}
   D(\phi) & C(\phi) \\
   -C(\phi) & -D(\phi) \\
  \end{array} } \right]
\end{equation}
This represents a system of linear differential equations with periodic variable coefficients. Its solution can be written as,
\begin{equation}
\boldsymbol{\Gamma}(\phi)=\boldsymbol{P}(\phi)e^{\phi  \boldsymbol{G}}
\end{equation}
where $\boldsymbol{P}(\phi)$ is a non-singular matrix, in this case with periodicity $\pi$, and   $\boldsymbol{G}$ is a constant matrix. 
The solution $\boldsymbol{\Gamma}(\phi)$ has the property that,
\begin{equation}
\boldsymbol{\Gamma}(\phi)=\boldsymbol{\Gamma}(\phi+\pi)=e^{\pi  \boldsymbol{G}}
\end{equation}
The eigenvalues of the matrix  $\exp^{\pi  \boldsymbol{G}}$, denoted as $\exp^{\pi \lambda_1}$ and $\exp^{\pi \lambda_2}$, are known as the quasienergies, while $\lambda_1$ and $\lambda_2$ are called the Floquet exponents). They satisfy the Liouville's formula,
\begin{equation}
\lambda_1+\lambda_2=\frac{1}{\pi}\int_0^{\pi} tr\boldsymbol{J}(\phi)d \phi
\end{equation}
which in this case leads to the condition $\lambda_1=-\lambda_2$. 

Unfortunately, there is no general method to further proceed and obtain $\boldsymbol{P}(\phi)$. Yet in this case we can reduce the system to a second order ordinary differential equation. To achieve such reduction, from Eq.(11) we  write, 
\begin{equation}
\Gamma_B(\phi)= \frac{1}{C(\phi)}\frac{d\Gamma_A}{d\phi} -\frac{D(\phi)}{C(\phi)}\Gamma_A, 
\end{equation}
as long as $C(\phi)\neq 0$. This already shows a fundamental difference between the pseudo scalar and pseudo vectorial field since $C(\phi)$ can be zero for a certain $\phi$ whenever $|\tilde{E}|<|\tilde{g}|$.

For the moment, assume that $C(\phi)\neq 0$. By taking the derivative of the previous equation,
\begin{equation}
\frac{d\Gamma_B(\phi)}{d\phi}= \frac{1}{C(\phi)}\frac{d^2\Gamma_A}{d\phi^2} +\frac{d}{d\phi}\left(\frac{1}{C(\phi)}\right)\frac{d\Gamma_A}{d\phi} -\frac{D(\phi)}{C(\phi)}\frac{d\Gamma_A}{d\phi}-\frac{d}{d\phi}\left(\frac{D(\phi)}{C(\phi)}\right)\Gamma_A, 
\end{equation}
we can introduce the two previous equations into Eq. (15), resulting in an uncoupled equation,

\begin{equation}
-\frac{1}{C(\phi)}\frac{d^2\Gamma_A}{d\phi^2} -\frac{d}{d\phi}\left(\frac{1}{C(\phi)}\right)\frac{d\Gamma_A}{d\phi} +\frac{D(\phi)}{C(\phi)}\frac{d\Gamma_A}{d\phi}+\frac{d}{d\phi}\left(\frac{D(\phi)}{C(\phi)}\right)\Gamma_A= C(\phi)\Gamma_A +\frac{D(\phi)}{C(\phi)}\frac{d\Gamma_A}{d\phi} -\frac{D^2(\phi)}{C(\phi)}\Gamma_A, 
\end{equation}
Collecting terms, the system of Eqns. (14) and (15) is transformed into an ordinary differential equation,
\begin{equation}\label{finalGamma}
-\frac{d^2\Gamma_A}{d\phi^2} +\frac{1}{C(\phi)}\frac{d C(\phi)}{d\phi}\frac{d\Gamma_A}{d\phi} +\left[C(\phi)\frac{d}{d\phi}\left(\frac{D(\phi)}{C(\phi)}\right)-C^2(\phi)  +D^2(\phi)\right]\Gamma_A=0, 
\end{equation}

The first derivative of $\Gamma_{A}(\phi)$ can be eliminated by using the following ansatz,
\begin{equation}
\Gamma_A(\phi)=\frac{Z(\phi)}{\sqrt{C(\phi)}}
\end{equation}
resulting that $Z(\phi)$ follows a Hill's equation,
\begin{equation}\label{Hill-s}
\frac{d^2 Z(\phi)}{d^2\phi}+F(\phi)Z(\phi)=0
\end{equation}
with $F(\phi)$ defined as,
\begin{equation}\label{Fofphi-s}
\begin{split}
F(\phi)&=-\left[D'(\phi)-\frac{C'(\phi)}{C(\phi)}D(\phi)-C^2(\phi)  +D^2(\phi)\right]+\\
&\frac{C''(\phi)}{2C(\phi)}-\frac{3}{4}\left(\frac{C'(\phi)}{C(\phi)}\right)^2
\end{split}
\end{equation}

\twocolumngrid

\bibliographystyle{unsrt}
\bibliography{refs.bib}

\begin{thebibliography}{10}

\bibitem{jiang2015review}
Jin-Wu Jiang, Bing-Shen Wang, Jian-Sheng Wang, and Harold~S Park.
\newblock A review on the flexural mode of graphene: lattice dynamics, thermal
  conduction, thermal expansion, elasticity and nanomechanical resonance.
\newblock {\em Journal of Physics: Condensed Matter}, 27(8):083001, 2015.

\bibitem{prb-jiang}
Jin-Wu Jiang, Hui Tang, Bing-Shen Wang, and Zhao-Bin Su.
\newblock Chiral symmetry analysis and rigid rotational invariance for the
  lattice dynamics of single-wall carbon nanotubes.
\newblock {\em Phys. Rev. B}, 73:235434, Jun 2006.

\bibitem{neto2009electronic}
AH~Castro Neto, F~Guinea, Nuno~MR Peres, Kostya~S Novoselov, and Andre~K Geim.
\newblock The electronic properties of graphene.
\newblock {\em Reviews of modern physics}, 81(1):109, 2009.

\bibitem{PRL-Vozmediano-92}
Jos\'e Gonz\'alez, Francisco Guinea, and M.~Angeles~H. Vozmediano.
\newblock Continuum approximation to fullerene molecules.
\newblock {\em Phys. Rev. Lett.}, 69:172--175, Jul 1992.

\bibitem{PRL-Kane-Mele-97}
C.~L. Kane and E.~J. Mele.
\newblock Size, shape, and low energy electronic structure of carbon nanotubes.
\newblock {\em Phys. Rev. Lett.}, 78:1932--1935, Mar 1997.

\bibitem{vozmediano2010gauge}
Mar{\'\i}a~AH Vozmediano, MI~Katsnelson, and Francisco Guinea.
\newblock Gauge fields in graphene.
\newblock {\em Physics Reports}, 496(4):109--148, 2010.

\bibitem{naumis-review}
Gerardo~G Naumis, Salvador Barraza-Lopez, Maurice Oliva-Leyva, and Humberto
  Terrones.
\newblock Electronic and optical properties of strained graphene and other
  strained 2d materials: a review.
\newblock {\em Reports on Progress in Physics}, 80(9):096501, 2017.

\bibitem{suzuura2002phonons}
Hidekatsu Suzuura and Tsuneya Ando.
\newblock Phonons and electron-phonon scattering in carbon nanotubes.
\newblock {\em Physical review B}, 65(23):235412, 2002.

\bibitem{Amorim}
B.~Amorim, A.~Cortijo, F.~de~Juan, A.G. Grushin, F.~Guinea,
  A.~Gutiérrez-Rubio, H.~Ochoa, V.~Parente, R.~Roldán, P.~San-Jose,
  J.~Schiefele, M.~Sturla, and M.A.H. Vozmediano.
\newblock Novel effects of strains in graphene and other two dimensional
  materials.
\newblock {\em Physics Reports}, 617:1 -- 54, 2016.
\newblock Novel effects of strains in graphene and other two dimensional
  materials.

\bibitem{Park-2014}
Cheol-Hwan Park, Nicola Bonini, Thibault Sohier, Georgy Samsonidze, Boris
  Kozinsky, Matteo Calandra, Francesco Mauri, and Nicola Marzari.
\newblock Electron–phonon interactions and the intrinsic electrical
  resistivity of graphene.
\newblock {\em Nano Letters}, 14(3):1113--1119, 2014.
\newblock PMID: 24524418.

\bibitem{Sohier-prb-2014}
Thibault Sohier, Matteo Calandra, Cheol-Hwan Park, Nicola Bonini, Nicola
  Marzari, and Francesco Mauri.
\newblock Phonon-limited resistivity of graphene by first-principles
  calculations: Electron-phonon interactions, strain-induced gauge field, and
  boltzmann equation.
\newblock {\em Phys. Rev. B}, 90:125414, Sep 2014.

\bibitem{barraza-prb}
James~V. Sloan, Alejandro A.~Pacheco Sanjuan, Zhengfei Wang, Cedric Horvath,
  and Salvador Barraza-Lopez.
\newblock Strain gauge fields for rippled graphene membranes under central
  mechanical load: An approach beyond first-order continuum elasticity.
\newblock {\em Phys. Rev. B}, 87:155436, Apr 2013.

\bibitem{levy2010strain}
N~Levy, SA~Burke, KL~Meaker, M~Panlasigui, A~Zettl, F~Guinea, AH~Castro Neto,
  and MF~Crommie.
\newblock Strain-induced pseudo--magnetic fields greater than 300 tesla in
  graphene nanobubbles.
\newblock {\em Science}, 329(5991):544--547, 2010.

\bibitem{ll-theory}
Francisco Guinea, MI~Katsnelson, and AK~Geim.
\newblock Energy gaps and a zero-field quantum hall effect in graphene by
  strain engineering.
\newblock {\em Nature Physics}, 6(1):30, 2010.

\bibitem{sasaki-pseudospin}
Ken-ichi Sasaki and Riichiro Saito.
\newblock Pseudospin and deformation-induced gauge field in graphene.
\newblock {\em Progress of Theoretical Physics Supplement}, 176:253--278, 2008.

\bibitem{georgi2017}
Alexander Georgi, Peter Nemes-Incze, Ramon Carrillo-Bastos, Daiara Faria,
  Silvia Viola~Kusminskiy, Dawei Zhai, Martin Schneider, Dinesh Subramaniam,
  Torge Mashoff, Nils~M Freitag, et~al.
\newblock Tuning the pseudospin polarization of graphene by a pseudomagnetic
  field.
\newblock {\em Nano letters}, 17(4):2240--2245, 2017.

\bibitem{anomalous}
ML~Ackerman, P~Kumar, M~Neek-Amal, PM~Thibado, FM~Peeters, and Surendra Singh.
\newblock Anomalous dynamical behavior of freestanding graphene membranes.
\newblock {\em Physical review letters}, 117(12):126801, 2016.

\bibitem{moving00}
YZ~He, Hui Li, PC~Si, YF~Li, HQ~Yu, XQ~Zhang, F~Ding, Kim~Meow Liew, and
  XF~Liu.
\newblock Dynamic ripples in single layer graphene.
\newblock {\em Applied physics letters}, 98(6):063101, 2011.

\bibitem{moving01}
EA~Korznikova and SV~Dmitriev.
\newblock Moving wrinklon in graphene nanoribbons.
\newblock {\em Journal of Physics D: Applied Physics}, 47(34):345307, 2014.

\bibitem{current01}
Tony Low, Yongjin Jiang, Mikhail Katsnelson, and Francisco Guinea.
\newblock Electron pumping in graphene mechanical resonators.
\newblock {\em Nano letters}, 12(2):850--854, 2012.

\bibitem{current02}
Abolhassan Vaezi, Nima Abedpour, Reza Asgari, Alberto Cortijo, and
  Mar{\'\i}a~AH Vozmediano.
\newblock Topological electric current from time-dependent elastic deformations
  in graphene.
\newblock {\em Physical Review B}, 88(12):125406, 2013.

\bibitem{current03}
Yongjin Jiang, Tony Low, Kai Chang, Mikhail~I Katsnelson, and Francisco Guinea.
\newblock Generation of pure bulk valley current in graphene.
\newblock {\em Physical review letters}, 110(4):046601, 2013.

\bibitem{current04}
Kai Zhang, Erhu Zhang, Huawei Chen, and Shengli Zhang.
\newblock Odd-parity currents induced by dynamic deformations in graphene-like
  systems.
\newblock {\em Journal of Physics: Condensed Matter}, 28(45):455301, 2016.

\bibitem{topo-naumis01}
Pedro Roman-Taboada and Gerardo~G Naumis.
\newblock Topological phase-diagram of time-periodically rippled zigzag
  graphene nanoribbons.
\newblock {\em Journal of Physics Communications}, 1(5):055023, 2017.

\bibitem{topo-naumis02}
Pedro Roman-Taboada and Gerardo~G Naumis.
\newblock Topological edge states on time-periodically strained armchair
  graphene nanoribbons.
\newblock {\em Physical Review B}, 96(15):155435, 2017.

\bibitem{topo-naumis03}
Pedro Roman-Taboada and Gerardo~G Naumis.
\newblock Topological flat bands in time-periodically driven uniaxial strained
  graphene nanoribbons.
\newblock {\em Physical Review B}, 95(11):115440, 2017.

\bibitem{topo-iade01}
Thomas Iadecola, David Campbell, Claudio Chamon, Chang-Yu Hou, Roman Jackiw,
  So-Young Pi, and Silvia~Viola Kusminskiy.
\newblock Materials design from nonequilibrium steady states: driven graphene
  as a tunable semiconductor with topological properties.
\newblock {\em Physical review letters}, 110(17):176603, 2013.

\bibitem{topo-iade02}
Thomas Iadecola, Titus Neupert, and Claudio Chamon.
\newblock Topological gaps without masses in driven graphene-like systems.
\newblock {\em Physical Review B}, 89(11):115425, 2014.

\bibitem{topological01}
Hannes Hübener, Umberto De~Giovannini, and Angel Rubio.
\newblock Phonon driven floquet matter.
\newblock {\em Nano Letters}, 0(0):null, 0.
\newblock PMID: 29361223.

\bibitem{tstatic01}
Mircea Trif, Pramey Upadhyaya, and Yaroslav Tserkovnyak.
\newblock Theory of electromechanical coupling in dynamical graphene.
\newblock {\em Physical Review B}, 88(24):245423, 2013.

\bibitem{tstatic02}
Ken-ichi Sasaki, Hideki Gotoh, and Yasuhiro Tokura.
\newblock Valley-antisymmetric potential in graphene under dynamical
  deformation.
\newblock {\em Physical Review B}, 90(20):205402, 2014.

\bibitem{moving02}
M~Oliva-Leyva and Gerardo~G Naumis.
\newblock Sound waves induce volkov-like states, band structure and collimation
  effect in graphene.
\newblock {\em Journal of Physics: Condensed Matter}, 28(2):025301, 2015.

\bibitem{moving03}
FJ~L{\'o}pez-Rodr{\'\i}guez and GG~Naumis.
\newblock Analytic solution for electrons and holes in graphene under
  electromagnetic waves: gap appearance and nonlinear effects.
\newblock {\em Physical Review B}, 78(20):201406, 2008.

\bibitem{moving04}
Richard Kerner, Gerardo~G Naumis, and Wilfrido~A G{\'o}mez-Arias.
\newblock Bending and flexural phonon scattering: Generalized dirac equation
  for an electron moving in curved graphene.
\newblock {\em Physica B: Condensed Matter}, 407(12):2002--2008, 2012.

\bibitem{electromechanical}
J~Scott Bunch, Arend~M Van Der~Zande, Scott~S Verbridge, Ian~W Frank, David~M
  Tanenbaum, Jeevak~M Parpia, Harold~G Craighead, and Paul~L McEuen.
\newblock Electromechanical resonators from graphene sheets.
\newblock {\em Science}, 315(5811):490--493, 2007.

\bibitem{nemes}
P{\'e}ter Nemes-Incze, Gerg{\H{o}} Kukucska, J{\'a}nos Koltai, Jen{\H{o}}
  K{\"u}rti, Chanyong Hwang, Levente Tapaszt{\'o}, and L{\'a}szl{\'o}~P
  Bir{\'o}.
\newblock Preparing local strain patterns in graphene by atomic force
  microscope based indentation.
\newblock {\em Scientific Reports}, 7(1):3035, 2017.

\bibitem{Klimov}
Nikolai~N. Klimov, Suyong Jung, Shuze Zhu, Teng Li, C.~Alan Wright, Santiago~D.
  Solares, David~B. Newell, Nikolai~B. Zhitenev, and Joseph~A. Stroscio.
\newblock Electromechanical properties of graphene drumheads.
\newblock {\em Science}, 336(6088):1557--1561, 2012.

\bibitem{Monteverde2015}
U.~Monteverde, J.~Pal, M.A. Migliorato, M.~Missous, U.~Bangert, R.~Zan,
  R.~Kashtiban, and D.~Powell.
\newblock Under pressure: Control of strain, phonons and bandgap opening in
  rippled graphene.
\newblock {\em Carbon}, 91:266 -- 274, 2015.

\bibitem{quantumwires}
Yong Wu, Dawei Zhai, Cheng Pan, Bin Cheng, Takashi Taniguchi, Kenji Watanabe,
  Nancy Sandler, and Marc Bockrath.
\newblock Quantum wires and waveguides formed in graphene by strain.
\newblock {\em Nano letters}, 18(1):64--69, 2017.

\bibitem{sasaki2008pseudospin}
Ken-ichi Sasaki and Riichiro Saito.
\newblock Pseudospin and deformation-induced gauge field in graphene.
\newblock {\em Progress of Theoretical Physics Supplement}, 176:253--278, 2008.

\bibitem{landau-elasticity}
Lev~Davidovich Landau and Eugin~M Lifshitz.
\newblock {\em Course of Theoretical Physics Vol 7: Theory and Elasticity}.
\newblock Pergamon Press, 1959.

\bibitem{Bai2014}
Ke-Ke Bai, Yu~Zhou, Hong Zheng, Lan Meng, Hailin Peng, Zhongfan Liu, Jia-Cai
  Nie, and Lin He.
\newblock Creating one-dimensional nanoscale periodic ripples in a continuous
  mosaic graphene monolayer.
\newblock {\em Phys. Rev. Lett.}, 113:086102, Aug 2014.

\bibitem{landau1959course}
Lev~Davidovich Landau and Eugin~M Lifshitz.
\newblock {\em Physique T\'eorique Vol 4: Electrodynamique quantique}.
\newblock Editions Mir, Moscow, 1989.

\bibitem{landau1960course}
Lev~Davidovich Landau and Eugin~M Lifshitz.
\newblock {\em Course of Theoretical Physics Vol 7: Quantum Electrodynamics}.
\newblock Pergamon Press, 1959.

\end{thebibliography}
\end{document}